# An Empirical Study on the Holiday Effect of China's Time-Honored Companies

Xianyang Li [a], Jiayi Xu [a], Haoxuan Xu [b], Yunxuan Ma [a], Yu Zhong [a], Lei Wang [a, *]

[a] School of Economics, Ocean University of China, Qingdao 266100, China

[b] School of Information Science and Engineering, Shandong University, Qingdao 266237, China

**Abstract:** The stock segment of China's time-honored brand enterprises has an important position in our securities stock market. The holiday effect is one of the market anomalies that occur in the securities market, which refers to the phenomenon that the stock market has significantly different returns than other trading days around festivals. The study of the holiday effect of China's time-honored brand enterprises can provide fresh ideas for the revitalization of our time-honored brands and the revitalization of time-honored enterprises. This paper takes listed companies of China's time-honored brand enterprises as the research object and focuses on the impact of the holiday effect on listed companies of China's time-honored brands with the help of the event study, and empirically analyses the changes in the return of listed companies of China time-honored brands during the Spring Festival period from 2012 to 2021. The empirical results reveal that: the time-honored brand concept stocks have a significant post-holiday effect during the Chinese New Year period, the time-honored alcoholic beverage enterprises are more sensitive to the Chinese New Year reflection, while the holiday effect of the time-honored pharmaceutical manufacturing enterprises is not significant.

**1. Introduction**

Spring Festival, which acts as a significant starting point in time for the Chinese nation and concentrates on the cultural traditions of China, Chinese national values, spirit, ethics and morals, habits of thinking, and aesthetic sensibilities in itself, is undoubtedly the grandest festival of the Chinese nation,

* Corresponding author. Lei Wang

E–mail address: wlei.123@163.com

the richest cultural connotation of all traditional festivals in China and an important carrier of the excellent Chinese traditional culture playing an important role in the inheritance of national culture and the protection of national culture[1]. The study of China's Time-honored Brand products is an important part of our nation's comprehensive culture[2], which is generally a memory product and service with a long history passed down from generation to generation. With a distinctive Chinese traditional cultural background, The China time-honored brand is often a collection of traditional culture and commercial culture integration. Considering the historical process, the China time-honored brand enterprises have an indispensable role in China's economy, and during the Spring Festival holiday, the China time-honored brand enterprises have an inescapable influence on the economy during the Spring Festival holiday.

However, since the beginning of the 21st century, the list of "Chinese time-honored" enterprises certified by the Ministry of Commerce totals 1128 enterprises, which is only 7% of the 16,000 time-honored enterprises at the beginning of the PCR[3]. Moreover, less than 10% of the existing 1,128 China time-honored brand enterprises are developing well, and the majority of them are experiencing business crises. Therefore, this study will focus on the impact of the Chinese New Year on the share price of China's time-honored brands, which will help revitalize their brands and thus provide us with a new guidance direction for revitalizing them. Revitalizing old Chinese brands not only protects the cultural heritage of the Chinese nation but also plays an exceedingly important role in promoting economic development, enhancing the competitiveness of the national economy, and expanding employment.

As there is no index for China's time-honored brand concept stocks in the market, it cannot be modeled using a time series model. This paper, therefore, uses Event Study to systematically analyze the stock returns oChina'sna time-honored brand-listed companies around the Chinese New Year from the perspective of industry analysis, using daily stock trading data for the ten-year period 2012-2021 as a sample to empirically analyze the Chinese New Year effect of China time-honored listed companies during this decade. Studying the phenomenon of abnormal returns of China's time-honored brand companies can provide a clear picture for China's time-honored brand companies, and investors and also help both to make better decisions.

This paper uses the event study to investigate the holiday effects of China's time-honored brand companies, filling the gaps in the festive effects of China's time-honored brand companies, followed by a heterogeneity test and robustness test for the old industry, and finally providing corresponding suggestions for the revitalization of old companies.

## 2. Literature Review

### 2.1 China Time-Honored Brand

Jiaxun H. E. chose a perspective combining intergenerational influence and brand equity, using an indigenous model of Chinese brand relationship quality (CBRQ), and found that old brand equity from older generations cannot be passed on to younger generations, except for trust[4]. Forêt, Philippe and et.al conducted a study on luxury goods that are Chinese time-honored brands as a sample and found that Chinese time-honored brands lack aesthetic differentiation and creativity, and argued that they should enhance their aesthetic management[5]. Zhang Y. conducted a study on time-honored companies in the Beijing area and found that most of these companies are not good at using self-media to promote their brands, and the marketing situation is not optimistic[6]. Price L.L, CoulterR. A. used a socio-historical-cultural approach and a psychological approach to deconstruct how cultural meanings are assembled into brands, pointing the way to the development of time-honored brands[7]. Yanni YAN, Tingting XIE et.al found that the accumulated know-how and heritage culture across generations help build strong brand equity, while entrepreneurship has an impact on these brands, brand equity, while entrepreneurship also has a positive impact on these firms[8]. Zhang S.N., Li Y.Q. et al. conducted a survey of 606 customers. The findings of the study highlighted that brand authenticity has a positive impact on consumers' in-person word-of-mouth and IWOM[9]. Song H, Xu J.B. et.al, who specified the study population as time-honored restaurants, found that nostalgia triggered by the food and service staff significantly evoked consumers' memories, while the dining environment stimulated the community component of the nostalgic experience, thus contributing to the development of the old-established economy[10]. Zhang H., Sun M. also studied the old-established restaurant industry in China. The study found that historical and traditional culture, regional transportation conditions, and urban development patterns all have important effects on the spatial distribution of old restaurant brands[11]. Song H., Kim J.H. found that brand heritage, directly and indirectly, influences purchase intention through brand authenticity and nostalgic experience[12]. Xu J.B., Prayag G. et.al used structural equation modeling and fsQCA to find that brand authenticity, brand image, and age all have a more significant effect on consumer loyalty towards old brands[13]. Fazal ur Rehman, Basheer M. Al-Ghazali found that brand image is significantly related to Malaysian consumers' purchase behavior towards fashion clothing brands by collecting questionnaires followed by data analysis and empirical testing[14]. Zhang P, Shi X and et.al using a cross-sectional regression model found that perceived fit has a significant impact on brand loyalty and also extension products with a high perceived fit which also affects brand loyalty to some extent[15]. After compiling the above literature, it is found that current research mostly studies the revitalization of old brands, brand culture, and purchase intention. The interaction between brand culture and festival culture has not been studied in depth, therefore, it is necessary to study whether old listed companies have "holiday effect".

## 2.2 Calendar Effect

Fields found that stocks have higher returns on the trading day before a specific religious holiday closure, and research on the holiday effect has proliferated since then[16]. Pettengill studied the S&P 500 from 1962-1986 and found higher pre-holiday returns for both large and small-sized firms, and smaller-sized firms had higher returns in both January and other months have higher returns than large-sized firms and argue that the holiday effect is due to the market closure[17]. Ariel finds that there is a significant pre-holiday effect in the U.S. stock market and also finds that the pre-holiday effect is not generated by other calendar effects (weekend effect and January effect.)[18]. Marrett G.J. and Worthington A.C. examine the holiday effect in the Australian stock market using a regression-based approach and find that Australian retail stocks have a significant pre-holiday effect but there is no post-holiday effect in any sector[19]. Urquhart A, and McGroarty F. use four calendar anomalies to study the market adaptation hypothesis and find that all four calendar anomalies support the market adaptation hypothesis and that the market adaptation hypothesis can explain the calendar anomalies well[20]. Hassan M.H. and Kayser M.S. using GJR-GARCH model on market return and trading volume data of the Dhaka Stock Exchange found that Ramadan has no significant relationship with the return and volatility of the Dhaka Stock Exchange but has a negative impact on the daily trade volume of DSE[21]. Ferrouhi E.M., Kharbouch O. and et.al using OLS regression with robust standard errors concluded that there is a Monday effect for BRVM and Namibia, a Friday effect for Kenya and Namibia, a January effect for Botswana and Zambia, a December effect, Ramadan effect for BRVM and Egypt, and Ramadan effect for Tunisia[22]. Hasan M.B., Hassan M.K. et.al used the GARCH model to conduct a comparative study on the holiday effect of conventional and Islamic stock indices in Bangladesh and found that both indices have calendar anomalies, with the differences being somewhat more pronounced in volatility[23]. Wuthisatian R. also used regression analysis to find the negative Monday effect and positive January effect in the Thai stock market[24]. Aslam F., Hunjra A. I., Tayachi T., et al studied eight Islamic frontier markets and the results of the study showed that Islamic frontier markets are weakly efficient markets whose investors do not consistently outperform the broad market when investing according to the Islamic or Gregorian calendar[25]. Since then research has shifted more towards studying the holiday effect in a particular sector. Qadan M., Aharon D.Y. et.al studied the calendar effect in the commodity market and found that the calendar effect is widely present in a representative set of nine natural resources, especially after the 21st century, the calendar effect in the commodity market has become more pronounced[26]. Chia-Ning Chiu began a study on the holiday effect in the restaurant industry by selecting data from selected restaurant firms in the U.S. from 2011-2016 and found that the majority of restaurant firms in the U.S. market have a significant holiday effect[27]. Kinateder H., and Papavassiliou V.G. turned the study to Bitcoin, using a GARCH-like model on the effect of holiday effects on Bitcoin daily conditional returns and volatility over the period 2013-2019, the results of the study showed no evidence to support the existence of an abnormal effect of Halloween on Bitcoin daily conditional returns[28]. Qadan M., Aharon D.Y. et.al found that the within-the-month effect is the only effect common to all cryptocurrencies[29]. Plastun A., Bouri E.

and et.al used statistical techniques (e.g., Student's t-test, ANOVA technique) to analyze the passion investment market. Statistical analysis found that there is a more significant Halloween effect in the passion investment market (mainly including tobacco, diamonds, etc.), but there is no year-end effect[30].

Through the calendar effect of the last five years, we found that the current research mainly focuses on a specific industry and a specific financial product, which is the hot spot of the current calendar effect research. However, no systematic research has been conducted on whether the "holiday effect" is prevalent in companies with distinctive brand cultures, such as long-established listed companies, and whether non-long-established companies have the holiday effect. Therefore, this paper uses the event research method to study whether the holiday effect exists in long-established companies.

## 3. Research Methods and Data Sources

*3.1 Research Methodology*

This section may be divided into subheadings. It should provide a concise and precise description of the experimental results, their interpretation, as well as the experimental conclusions that can be drawn.

This paper focuses on the "holiday effect" of long-established listed companies using the event study method. The event study method is often used to study abnormal stock price returns when a particular event occurs[31,32]. The main principle is that a particular event is selected for a specific purpose and stock return changes are studied for a specific length of time before and after the event in order to assess the impact of the event on stock returns. The specific steps include.

(1) Defining events and time windows. The estimation window, the event window and the post-event window are important components of the event study method. As in Figure 1, $T = 0$ is the event date, T to $T_0$ is the estimation window, $T_1$ to $T_2$ is the event window. $T_2$ to $T_3$ is the ex-post window. There is no uniform guideline for the choice of event window, but it must be based on preventing overlap between policies and other events[33,34]. Specifically for this paper, the five days before and after Chinese New Year are the event window (i.e. T-5-T+5). The 100 trading days before the event window is the estimation window, i.e. T-105-T-6.

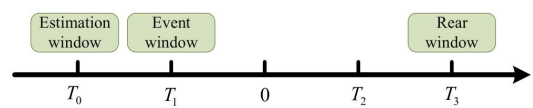

**Figure 1.** Time windows of event study

(2) Selection of the research sample. This paper uses the representative Chinese New Year in a year as the research event and selects stock trading data for the ten-year period 2012-2021 as the research sample. A total of 37 stocks were selected from listed companies whose trademarks were included in the Ministry of Commerce's Chinese China time-honored brand List and listed earlier than 2011, including the restaurant sector (Quanjude, Xi'an Catering, Shanghai Meilin, Hengshun Vinegar, Bright Dairy), the wine, beverage, and refined tea manufacturing sector (Guizhou Maotai, Laobai Ganjiu, Shanxi Fenjiu, Luzhou Laojiao, Wuliangye, Shunxin Agriculture, Yanghe Co. Guoyue Longshan, Jinfeng Liquor, Tsingtao Beer, Zhang Yu A, Hainan Coconut Island, Tuopai Shede, Shuijingfang), Pharmaceutical Manufacturing Sector (Shanghai Pharmaceutical, Zhejiang Zhenyuan, Tongrentang, Baiyunshan, Pientzehuang, Jianmin Group, Ma Yinglong, Zhongxin Pharmaceutical, Yunnan Baiyao, Ha Pharmaceutical, Jiuzitang, Guangyuyuan), Retail & Wholesale Sector (Kaikai Industrial, Yuyuan, New World, Wangfujing), Other Sectors (Laofengxiang, Shanghai Phoenix).

(3) Selection of the normal return model. In this paper, the market model is chosen as the estimation model for normal returns in conjunction with the actual situation of the research subject, so that the return on a particular security can be combined with the market return.

$$R_{it} = \alpha_i + \beta_i I_t + \varepsilon_{it} \tag{1}$$

$$E(\varepsilon_{it}) = 0 \tag{2}$$

$$Var(\varepsilon_{it}) = \sigma_{it}^2 \quad \varepsilon_i \in (0, \sigma_{it}^2) \tag{3}$$

Where $R_{it}$ is the return on listed company stocks; $I_t$ is the market return, and the market return of SSE Composite Index and SZSE Composite Index are selected in this paper.

(4) Determine the model for estimating abnormal returns and cumulative abnormal returns. In the event study approach, abnormal returns are calculated using the formula $AR_{it} = R_{it} - ER_{it}$. In this equation, the $ER_{it}$ is calculated as determined by the normal rate of return model above, the $R_{it}$ is the actual return for that stock.

In that case, we calculate the average excess return of the sample based on a given $N$ sample of events, separately calculate the sample average excess return for the sample $AAR_t$, and the cumulative excess return $CAR(t_1, t_2)$, $T_2 < t_1 < t_2 < T_3$.

$$AAR_t = \frac{1}{N}\sum_{i=1}^{N} AR_{it}$$
(4)

$$CAR(t_1, t_2) = \sum_{t=t_1}^{t_2} AR_{it}$$
(5)

(5) Testing the significance of abnormal returns. Considering the number of samples, the t-test method is more appropriate for this study to test the significance of the average cumulative abnormal returns.

$$T(i) = \frac{CAR(i)}{\sqrt{\frac{T}{N}}}$$
(6)

where N is the number of all samples and $i$ is the number of $i$ sample.

*3.2 Selection of Variables and Processing of Data*

A total of 60 China time-honored enterprises have been collected. In order to better reflect the development status of China's time-honored listed companies and to select more representative mature listed companies with distinctive time-honored characteristics, the criteria of the study sample selection are as follows:

(1) Companies that are not on the Ministry of Commerce's Chinese China time-honored brand List, six companies: Foshan Lighting (000541), Great Eastern (600327), Sanyuan (600429), Gujing Gongjiu (000596), Fosun Pharmaceutical (600196), Jiayue (603708).

(2) Excluding stocks listed after 2011 (i.e. less than 10 years old), including Zhang Xiaoquan (301055) listed in 2021, Tongqinglou (605108) listed in 2020, Guangzhou Restaurant (603043) listed in 2017, Guifaxiang (002820) listed in 2016, Yingjia Gongjiu (603198) and Kouzijiao (603589) listed in 2015, Haitian Flavor (603288)、Jinshiyuan (603369) and Huijishan (601579) listed in 2014.

(3) Also added to the list of companies recognized by local governments as long-established local companies: Hainan Coconut Island (600238), Shuijingfang (600779).

Ultimately, this paper selected 37 long-established listed companies that met the criteria, all of which had the fundamental characteristics which are representative of long-established listed companies. The data were obtained from the Guotaian database and processed using STATA statistical software.

**4. Empirical Analysis of the "Holiday Effect" in the Stocks of Long-Established Companies**

*4.1 Pre-Epidemic.*

4.1.1 The average return and average cumulative return of the holiday effect

of time-honored stocks before the epidemic

The average abnormal returns and average cumulative abnormal returns of the 37 long-established listed companies over the eight Chinese New Year events from 2012 to 2019 are shown in Table 1.

**Table 1** Average Excess Return (AAR) and Average Cumulative Excess Return (CAR) over the 2012-2019 Chinese New Year Event Period.

| Event Day | 2016 AAR | 2016 CAR | 2017 AAR | 2017 CAR | 2018 AAR | 2018 CAR | 2019 AAR | 2019 CAR |
|---|---|---|---|---|---|---|---|---|
| -5 | 0.826 | 0.826 | -0.067 | -0.067 | 2.746 | 2.746 | 0.228 | 0.228 |
| -4 | 0.168 | 0.994 | 0.062 | -0.005 | 1.562 | 4.309 | -0.034 | 0.194 |
| -3 | -0.451 | 0.543 | 0.097 | 0.091 | 1.123 | 5.432 | -1.141 | -0.947 |
| -2 | -0.062 | 0.481 | -0.079 | 0.013 | -0.588 | 4.844 | -0.060 | -1.006 |
| -1 | -0.387 | 0.094 | -0.141 | -0.128 | 0.114 | 4.958 | 0.340 | -0.666 |
| 0 | 0.476 | 0.570 | 0.595 | 0.466 | -0.449 | 4.508 | 0.968 | 0.302 |
| 1 | -1.262 | -0.692 | 0.435 | 0.901 | -0.519 | 3.990 | 0.486 | 0.788 |
| 2 | -1.054 | -1.746 | 0.493 | 1.393 | -0.480 | 3.510 | -1.028 | -0.241 |
| 3 | 0.223 | -1.523 | -0.112 | 1.282 | 0.692 | 4.202 | 1.037 | 0.796 |
| 4 | -0.251 | -1.775 | 0.640 | 1.922 | -0.378 | 3.824 | 0.399 | 1.195 |
| 5 | -0.774 | -2.549 | -0.241 | 1.680 | 0.150 | 3.974 | -0.059 | 1.136 |

| Event Day | 2012 AAR | 2012 CAR | 2013 AAR | 2013 CAR | 2014 AAR | 2014 CAR | 2015 AAR | 2015 CAR |
|---|---|---|---|---|---|---|---|---|
| -5 | -1.891 | -1.891 | -1.089 | -1.089 | 0.020 | 0.020 | 0.382 | 0.382 |
| -4 | 0.190 | -1.701 | 1.984 | 0.895 | 0.208 | 0.228 | 0.397 | 0.778 |
| -3 | -1.311 | -3.012 | 0.678 | 1.573 | -0.444 | -0.216 | 0.510 | 1.288 |
| -2 | 0.441 | -2.571 | 1.153 | 2.726 | -1.411 | -1.627 | 0.636 | 1.924 |
| -1 | 1.379 | -1.192 | 1.512 | 4.238 | -0.174 | -1.801 | 0.043 | 1.966 |
| 0 | 1.140 | -0.052 | -0.020 | 4.218 | 0.070 | -1.731 | 0.365 | 2.332 |
| 1 | 0.289 | 0.237 | 1.814 | 6.032 | 0.110 | -1.621 | 1.238 | 3.570 |
| 2 | 0.710 | 0.947 | 2.069 | 8.100 | 0.770 | -0.851 | 1.403 | 4.973 |
| 3 | -0.503 | 0.445 | 2.187 | 10.287 | 0.755 | -0.097 | 0.992 | 5.964 |
| 4 | 0.218 | 0.663 | 1.712 | 11.999 | 0.285 | 0.188 | 0.220 | 6.185 |
| 5 | 0.777 | 1.440 | -0.570 | 11.429 | 0.919 | 1.107 | 1.772 | 7.957 |

Using the event study methodology and the selected model, the average daily excess return for all stocks in the sample was calculated $AAR_t$ and the cumulative average daily excess return $CAAR_t$. To better illustrate the empirical results of the analysis of the sample data, the paper is illustrated in the form of charts. Figures 2 to 11 show the average daily excess return and cumulative average daily excess return of all stocks in the sample for the Chinese New Year period from 2012 to 2021. The graphs of $AAR_t$ and cumulative average daily excess return $CAAR_t$ trends can be classified into four categories.

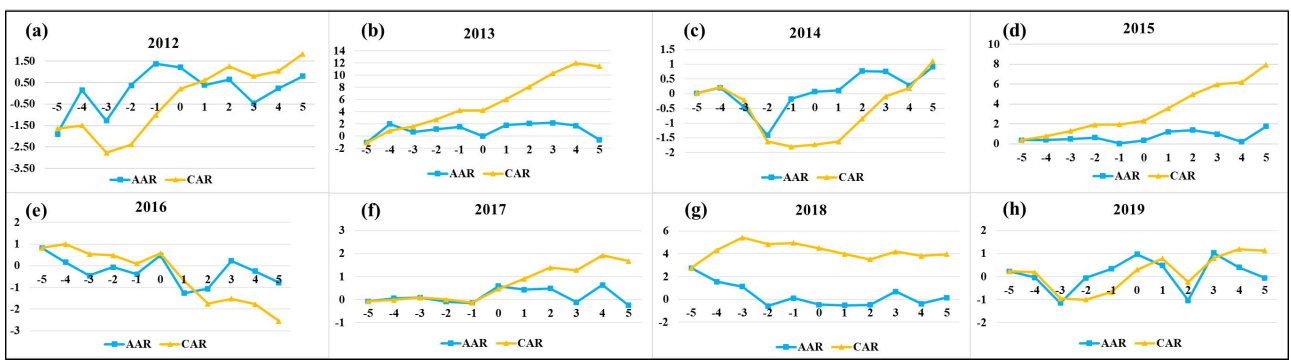

**Figure 2.** 2012-2019 Spring Festival abnormal income and cumulative abnormal income trend

Figure 2 (a、f、h) show the first category. 2012 (Figure 2a) the average cumulative excess return (CAR) for the 5 trading days prior to the event date before the Chinese New Year is below zero, while the average cumulative excess return (CAR) for the trading days after the event date is positive. 2017 (Figure 2f) the 1、4 and 5th days before the event date are all below zero, while after the event date, they are all above zero and show an increasing trend. In 2019 (Figure 9), the average cumulative excess return (CAR) was negative on 3 of the 5 trading days prior to the event date and positive on all but the second day after the holiday. The above six charts of excess returns and average cumulative excess returns illustrated a positive pre-holiday effect and a positive post-holiday effect for the oldest listed companies in 2012, 2017 & 2019.

Figure 2 (b、d、g) show the second category. 2013 (Figure 2b) has a positive average cumulative excess return, except for the 5th day before the event date, which is negative. 2015 (Figure 2d) and 2018 (Figure 2g) both show a volatile growth trend with a positive average cumulative excess return (CAR) throughout the event window. The images for these three years demonstrated how the arrival of the Chinese New Year contributed to the growth of the old economy in 2013, 2015, and 2018.

Figure 2 (c) shows the third category. the average cumulative excess return (CAR) for 2014 (Figure 2c) was negative for both the three days before and the three days after the event day, and the average cumulative excess return was below zero for six of the ten trading days in the event window. The above illustrated that in 2014, the arrival of the Chinese New Year had a negative effect on China's time-honored brand companies.

Figure 2 (e) shows the fourth category. 2016 (Figure 2e) has a positive average cumulative excess return (CAR) for all five trading days prior to the event date and a negative average cumulative excess return (CAR) for all five trading days after the event date, with the image showing a volatile downward trend overall. This suggests a more pronounced positive pre-holiday and negative post-holiday effect for China's time-honored brand companies in 2016. And in relation to the stock market crash in China at the beginning of 2016. On 4 January (the first day of trading) and 7 January 2016, the Shanghai and Shenzhen stock markets collapsed and triggered a new meltdown mechanism. In addition, 2016 saw severe turmoil in overseas financial markets, with stock markets plunging en masse from the Americas to Europe to Asia. As a result, after the festive period, investors were generally not bullish on the mainland stock market due to the downward trend before the Chinese New Year and the general plunge in both domestic and overseas stock markets. These black swan events made the actual returns of the China time-honored concept stocks smaller than their actual returns, thus making the China time-honored listed companies present a negative post-holiday effect in 2016.

In summary, over the eight-year period from 2012-2019, the holiday effects were not identical across the Chinese New Year. Seven of these Spring Festivals showed positive abnormal returns, with six of them showing positive post-holiday returns, proving that the event of Chinese New Year drives the old economy after the festival.

In order to further explore the holiday effect of China's time-honored brand companies, this paper further integrates the $AAR_t$ for the eight Chinese New Year samples from 2012-2019 and finds its mean values $AAAR_t$ and $CAAAR_t$, within twenty trading days of the event window, by the following formula.

$$AAAR_t = \frac{1}{8}\sum_{n=1}^{8} AAR_t \quad (7)$$

$$CAAR_t = \frac{1}{8}\sum_{t=t_1}^{t_2} AAAR_t \quad (8)$$

The two variables mentioned above are referred to as the combined average daily abnormal return and the combined cumulative average daily abnormal return. Based on the results of the calculations, we plotted the trend of the two as follows:

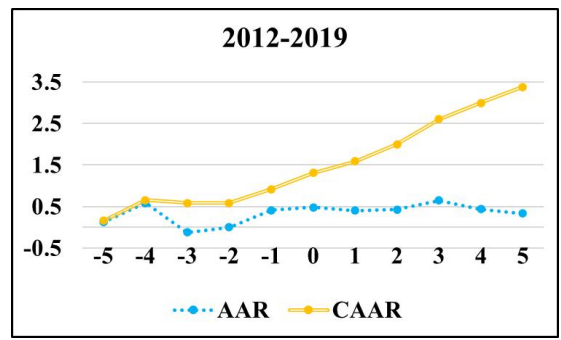

**Figure 3.** The trend of Average Abnormal Returns and Average Cumulative Abnormal Returns in Spring 2012-2019

The chart above shows the trend of $AAAR_t$ and $CAAAR_t$ prior to the outbreak. We can see that prior to the event trading day, the $CAAAR_t$ was above zero on all event days, although the $AAAR_t$ was below zero on the 3rd day before the event day, and showed a fluctuating growth trend, especially after the holiday season when the growth trend intensified and became more pronounced. This result shows that the Chinese New Year has a positive impact on the China time-honored concept stocks, and the positive impact is more significant after the festival, which is in line with the results of the previous analysis of the 10 charts. Based on this, we can conclude that the traditional Chinese New Year festival has a short-term positive impact on the China time-honored economy after the festival, contributing to the growth of the China time-honored economy after the festival.

4.1.2 Analysis of the significance of the average cumulative abnormal return of the holiday effect of time-honored stocks before the epidemic

During the period from 2012 to 2019, the average cumulative abnormal returns of China's time-honored listed enterprises and their significance test results are shown in Table 2, and the time intervals [-5, -1], [-2,2], [1,5] indicate before, during and after the Spring Festival, respectively. The empirical results show that in the period before the Chinese New Year (-5 to 1 day), all except 2016, 2017, and 2019 show a significant holiday effect, with 2012 and 2014 showing a significant negative effect. Among the Chinese New Year events (-2-2 days), 2012, 2013, 2015, 2016, and 2018 all showed significant holiday effects, with 2016 & 2018 showing negative significant effects. In the post-Chinese New Year (1-5 days) 5 out of 8 Chinese New Year showed a significant post-holiday effect and the significance level reached 1% in 2013, 2015, and 2016, with 2016 showing a negative significance. The results of the significance test found that the Chinese New Year had a significant effect on the China's time-honored concept stocks, with the positive post-holiday effect being more significant. The empirical results of this paper are in line with the theoretical inference of the impact of the Chinese New Year China's time-honored listed companies.

**Table 2.** Results of the Mean Cumulative Abnormal Returns Test for the Chinese New Year effect,

2012-2019

| Time interval | 2012 | 2013 | 2014 | 2015 | 2016 | 2017 | 2018 | 2019 |
|---|---|---|---|---|---|---|---|---|
| [-5,-1] | -1.192* | 4.238*** | -1.801* | 1.966** | 0.094 | -0.128 | 4.958*** | -0.666 |
| [-2,2] | 3.959*** | 6.527*** | -0.635 | 3.685** | -2.289** | 1.302 | -1.921*** | 0.706 |
| [1,5] | 1.492* | 7.212*** | 2.838** | 5.625*** | -3.119 *** | 1.214 | -0.535 | 0.835 |

Note: *, **, *** denote 10%, 5%, 1% significance levels respectively.

*4.2 After the Outbreak.*

4.2.1 The average return and average cumulative return of the holiday effect of time-honored stocks after the epidemic

Table 3 show the average daily excess return AARt and the average cumulative daily excess return CARt for 2020 and 2021 for established listed companies. To facilitate the analysis, a line chart as shown in Figure 4 was drawn. As can be seen from Figure 4(a), 2020 has only the first and fifth trading days after the event date in the event window that are less than zero, and eight of the ten trading days in the event window are positive. 2021 (Figure 4b) has all 10 trading days in the event window with a CAR below zero. The average cumulative excess return (CAR) for all 10 trading days in the event window in 2021 is below zero. This suggests that in 2020 the arrival of the Chinese New Year boosts the old economy at that time, while in 2021 the arrival of the Chinese New Year shows some negative impact on the old economy.

**Table 3.** Results of the Mean Cumulative Abnormal Returns Test for the Chinese New Year effect, 2020-2021

| Event Day | 2020 | | 2021 | |
|---|---|---|---|---|
| | AAR | CAR | AAR | CAR |
| -5 | 0.527 | 0.527 | -0.255 | -0.255 |
| -4 | 0.536 | 1.064 | 0.253 | -0.002 |
| -3 | 0.776 | 1.839 | -0.263 | -0.266 |
| -2 | -1.478 | 0.361 | -0.058 | -0.324 |
| -1 | -0.291 | 0.070 | -0.162 | -0.486 |
| 0 | -0.338 | -0.267 | -1.292 | -1.778 |
| 1 | -0.789 | -1.056 | 0.569 | -1.209 |
| 2 | 1.618 | 0.561 | -1.182 | -2.391 |
| 3 | 0.037 | 0.598 | 0.137 | -2.253 |
| 4 | -0.141 | 0.457 | -0.621 | -2.874 |
| 5 | -0.996 | -0.539 | -1.442 | -4.316 |

**Figure 4.** 2020–2021 Chinese New Year Anomaly Earnings and Cumulative Abnormal Return Trends

### 4.2.2 Analysis of the significance of the average cumulative abnormal return of the holiday effect of time-honored stocks after the epidemic

Table 4 show the average cumulative excess returns and their significance test results for the China time-honored brand companies in 2020 and 2021. Whether before, during, or after Chinese New Year, the effect of Chinese New Year on the China time-honored brand companies is not significant, except for 2021 when Chinese New Year shows a negative significance level of 10%.

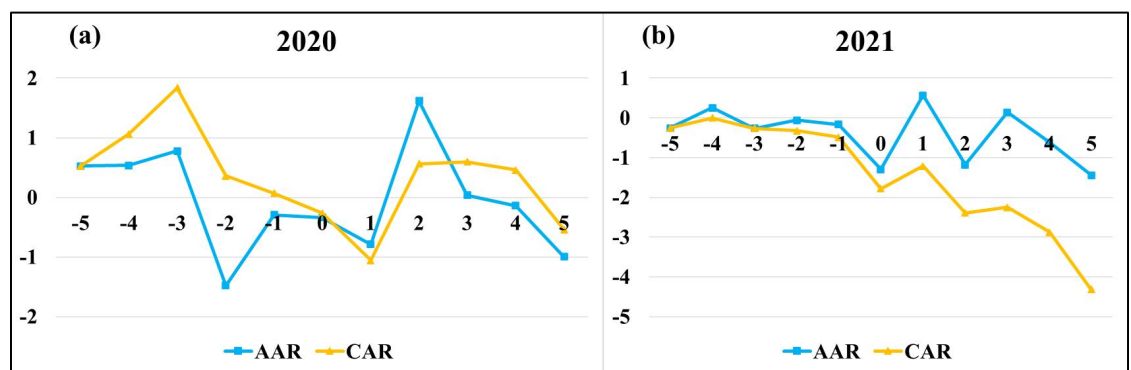

The possible reason for this situation is that, after the epidemic, the pharmaceutical sector of China's time-honored concept stocks was more significantly affected by the epidemic and was favored by investors. However, many of these sectors, such as alcohol and food and beverage, were affected by the epidemic and many of their businesses were stagnant and not favored by investors, resulting in lower profit margins and therefore some decline. The market, therefore, did not react significantly to the combined negative and positive effects of the epidemic.

**Table 4.** Results of the average cumulative abnormal returns test for the Chinese New Year effect in 2020-2021

| Time interval | 2020 | 2021 |
|---|---|---|
| [-5,-1] | 0.070 | -0.486 |
| [-2,2] | -1.278 | -2.125 |
| [1,5] | -0.272 | -2.538 * |

Note: *, **, *** denote 10%, 5%, 1% significance levels respectively.

## 5. Heterogeneity Test

### 5.1 Industry Heterogeneity

According to the 2012 edition of the SFC's industry classification, established listed enterprises can be broadly classified into four categories - wine, beverage and refined tea manufacturing, medical manufacturing, wholesale and retail, and food and beverage, with wine, beverage and refined tea manufacturing and medical manufacturing accounting for a relatively large proportion. The non-China time-honored enterprises in these two industries are now subjected to segmental significance tests and compared with the China time-honored enterprises.

5.1.1 Heterogeneity in the Alcohol, Beverage and Refined Tea Manufacturing Sector

The graph 5 shows that in 6 out of 10 years, there is a significant holiday effect of varying degrees for established companies in the wine, beverages and refined tea manufacturing industry. In contrast, non-established companies in the wine, beverage and refined tea manufacturing industry only showed positive festival effects in 2012 and 2013, and negative festival effects in 2015 and 2020, with only six out of 10 years showing a low level of significance. Overall, the Chinese New Year event has a greater impact on the share prices of long-established companies in the wine, beverages and refined tea manufacturing sector than it does on the non-long-established companies.

Tests for between-group differences in heterogeneity in alcohol, beverage and refined tea manufacturing.

The demand for alcohol during the Chinese New Year is huge, especially since high-end liquor is in short supply, unable to meet consumer demand, and has a large room for price increases. The time-honored alcoholic beverage companies, with their unique craftsmanship and sophisticated products, as well as the added benefit of their historical origins, occupy the majority of the high-end liquor market. Therefore, during the Chinese New Year, the old liquor companies have a more significant holiday effect, especially a more significant positive effect after the festival.

**Table 5.** Results of the average cumulative abnormal returns test for the Chinese New Year effect in the wine, beverages and refined tea manufacturing sector, 2012-2021

| Event Window | 2012 | | 2013 | | 2014 | | 2015 | | 2016 | |
|---|---|---|---|---|---|---|---|---|---|---|
| | Long-established | Non-Elderly | Long-established | Non-Elderly | Long-established | Non-Elderly | Long-established | Non-Elderly | Long-established | Non-Elderly |
| [-5,-1] | -1.740 | -1.747 | 2.501** | 0.962 | -1.296 | -2.508 | 2.281 | -1.663 | -1.269* | -1.616 |
| [-2,2] | 2.492** | 2.870 | 5.591*** | 7.012*** | 0.098 | -1.688 | 4.925 | -0.449 | -2.426* | 0.984 |
| [1,5] | 1.325 | 2.755*** | 7.262*** | 10.194*** | 4.346* | 0.773 | 8.441** | 1.358 | -1.903 | 1.124 |

| Event Day | 2017 | | 2018 | | 2019 | | 2020 | | 2021 | |
|---|---|---|---|---|---|---|---|---|---|---|
| | Long-established | Non-Elderly | Long-established | Non-Elderly | Long-established | Non-Elderly | Long-established | Non-Elderly | Long-established | Non-Elderly |
| [-5,-1] | -0.593 | -0.790 | 2.378** | 1.162 | 0.227 | -0.599 | 0.265 | -3.694 | 0.624 | -4.567 |
| [-2,2] | -0.633 | 0.756 | -0.699 | -2.157 | 1.167 | 0.875 | -1.630 | -7.831*** | -2.611 | 0.656 |
| [1,5] | 0.566 | -0.845 | -2.395 | 2.637 | -0.656 | 0.981 | -2.890 | -1.282*** | -0.660 | -3.867 |

Note: *, **, *** denote 10%, 5%, 1% significance levels respectively.

*5.1.2 Pharmaceutical Manufacturing Heterogeneity*

In the SFC Class of 2012 industry classification, 99 non-time-honored pharmaceutical manufacturing companies were finally selected and tested for significance by interval, taking into account factors such as time on market and operating status. Table 6 shows a graph comparing the holiday effect between the old pharmaceutical manufacturing industry and the non-old pharmaceutical manufacturing listed companies.

The graph 6 shows that non-China time-honored brand companies in the pharmaceutical manufacturing category show varying levels of the holiday effect in all nine years from 2012-2021 except 2018. Among them, 2012, 2013, 2015 and 2019 showed significant festive effects in the event window. In 2012, the significance level of the pre-holiday effect of non-established listed enterprises decreased from 5% to 1%, while the significance level during and after the Spring Festival showed a positive post-holiday effect of 10%. In 2013, they show a holiday effect at a significance level of 5% before the event date, while they also show a holiday effect at a significance level of 10% during (-2~2 days) and after (1~5 days) the Chinese New Year. The positive post-holiday effect with a significance level of 10% is presented in 2015 and 2019 during the pre-holiday period, during Chinese New Year, and after Chinese New Year. The remaining years also show varying levels of significance in the corresponding intervals of the event window. In contrast, non-established pharmaceutical manufacturing listed companies only had varying degrees of festive effects in six of the past ten years, with only 2013 showing significant festive effects in the entire day of the event, with significantly lower numbers and significantly lower levels of significance than non-established pharmaceutical manufacturing companies. Overall, the Chinese New Year had a significantly greater impact on non-established pharmaceutical manufacturing companies than established pharmaceutical manufacturing companies, i.e. non-established pharmaceutical manufacturing companies had a more significant holiday effect during the Chinese New Year.

Pharmaceutical manufacturing class enterprises except Pientzehuang, Yunnan Baiyao, and other enterprises, the annualized return of the rest of the stocks is only 5%, and the whole Chinese medicine concept stocks are not innovative enough, the inward volume is serious, the whole industry is in shrinkage, investors think that the investment value of the sector is not high, and do not look forward to the old pharmaceutical manufacturing class concept stocks. But the old pharmaceutical manufacturing class listed companies will still make some unconventional profits under the role of brand culture and Chinese New Year culture. Therefore, the market response of the China time-honored pharmaceutical manufacturing category companies under the combined effect of positive and negative factors is insignificant, presenting and less festive effect than the non-China time-honored manufacturing category listed companies.

**Table 6.** Results of the average cumulative abnormal returns test for the Chinese New Year effect in the Pharmaceutical manufacturing sector, 2012-2021

| Event Window | 2012 | | 2013 | | 2014 | | 2015 | | 2016 | |
|---|---|---|---|---|---|---|---|---|---|---|
| | Long-established | Non-Elderly | Long-established | Non-Elderly | Long-established | Non-Elderly | Long-established | Non-Elderly | Long-established | Non-Elderly |
| [-5,-1] | -1.781** | -1.077* | 2.357** | 1.100** | -1.095 | 1.106** | 1.724 | 1.349*** | -1.149* | 0.396 |
| [-2,2] | 2.594** | 2.757*** | 4.975** | 5.514*** | -0.025 | 0.727 | 2.969 | 2.949*** | -2.236 | 0.344 |
| [1,5] | 1.248 | 1.411*** | 7.524*** | 3.576*** | 4.242 | 1.976*** | 7.909* | 5.695*** | -1.021 | -2.221*** |

| Event Window | 2017 | | 2018 | | 2019 | | 2020 | | 2021 | |
|---|---|---|---|---|---|---|---|---|---|---|
| | Long-established | Non-Elderly | Long-established | Non-Elderly | Long-established | Non-Elderly | Long-established | Non-Elderly | Long-established | Non-Elderly |
| [-5,-1] | -0.768 | 0.432 | 2.549** | 1.098 | 0.188 | -1.205*** | -0.034 | 8.754*** | 1.798 | 0.709 |
| [-2,2] | -0.767 | 0.502 | -0.659 | -0.997 | 1.038 | 1.207*** | -2.638 | 13.339*** | -3.996* | 5.953*** |
| [1,5] | 0.961 | -0.794** | -2.464 | 0.445 | -0.990 | 1.938*** | -2.799 | 0.679*** | -1.793 | 5.388*** |

Note: *, **, *** denote 10%, 5%, 1% significance levels respectively.

**Table 7.** Results of the mean cumulative abnormal returns test for the Chinese New Year effect with an estimation window of 50 days, 2012-2019

| Time interval | 2012 | 2013 | 2014 | 2015 | 2016 | 2017 | 2018 | 2019 |
|---|---|---|---|---|---|---|---|---|
| [-5,-1] | -1.565** | 4.285*** | -1.934** | 1.986** | -0.133 | -0.042 | 5.272*** | -1.101* |
| [-2,2] | 1.183 | 7.266*** | 2.321* | 5.519*** | -3.199*** | 1.300 | -0.046 | 0.276 |
| [1,5] | 3.749*** | 6.579*** | -1.112 | 3.668** | -2.398** | 1.332 | -1.215** | 0.067 |

Note: *, **, *** denote 10%, 5%, 1% significance levels respectively.

## 6. Robustness Tests

This section is not mandatory but may be added if there are patents resulting from the work reported in this manuscript.

To verify the robustness of the results, different estimation window lengths used for market model estimation were used. Given that the number of trading days between each event day is approximately 200 days, and that Degryse[35] argues that the estimation window should take a range of values from 50 to 1000, we choose a 50-day estimation window for estimation because of the large difference in length between a 50-day estimation window and a 100-day estimation window. Table 7 reports the average cumulative abnormal returns over the first eight events of the epidemic and the results of their tests for an estimation window of 100 days. Therefore, we can conclude that the significance test results in Tables 2 and 4 are robust.

Pharmaceutical manufacturing class enterprises except Pientzehuang, Yunnan Baiyao and other enterprises,

**Table 8.** Results of the average cumulative abnormal returns test for the Chinese New Year effect with a 50-day estimation window for 2020-2021

| Time interval | 2020 | 2021 |
|---|---|---|
| [-5,-1] | -0.082 | -0.258 |
| [-2,2] | -0.239 | -2.946* |
| [1,5] | -1.488 | -1.934 |

Note: *, **, *** denote 10%, 5%, 1% significance levels respectively.

using a 50-day estimation window. The test results show that there is a significant holiday effect for seven of the eight events. Of these, five of the eight event days after the Chinese New Year (1-5) show significant post-holiday effects, with 2016 and 2018 showing negative significance.

Table 8 reports the average cumulative abnormal returns for the two events following an epidemic and the results of their tests. The results of the tests show that with an estimation window of 50 days, the China time-honored brand firms after the outbreak do not show a significant holiday effect.

As shown in Tables 7 and 8, the significance test results for an estimation window of 50 days are similar to those the annualized return of the rest of the stocks is only 5%, and the whole Chinese medicine concept stocks are not innovative enough, the inward volume is serious, the whole industry is in shrinkage, investors think that the investment value of the sector is not high, and do not look forward to the old pharmaceutical manufacturing class concept stocks. But the old pharmaceutical manufacturing class listed companies will still make some unconventional profits under the role of brand culture and Chinese New Year culture. Therefore, the market response of the China time-honored pharmaceutical manufacturing category companies under the combined effect of positive and negative factors is insignificant, presenting and less festive effect than the non-China time-honored manufacturing category listed companies.

## 7. Conclusions and analysis

Using the event study method, this paper divides the period 2012-2021 into two parts, before and after the outbreak, to systematically study the market response to the "festive effect" during the Chinese New Year period for long-established listed companies. The empirical results show that in 2012-2019, the China time-honored brand companies showed a more significant positive post-epidemic effect. In the post-epidemic years of 2020 and 2021, China time-honored brand companies do not show a significant holiday effect due to the epidemic. At the same time, we conducted a heterogeneity analysis after the sectoral breakdown and found that listed companies in the category of China time-honored wine, beverage and refined tea manufacturing had a more significant festive effect compared to non-China time-honored wine, beverage, and refined tea manufacturing companies, while listed companies in the category of non-China time-honored pharmaceutical manufacturing had a more significant festive effect compared to China time-honored pharmaceutical manufacturing companies due to composition factors.

## 8. Recommendations for the response

During the holiday season, people tend to take the opportunity to enjoy themselves and relax, so shopping becomes the obvious choice. Businesses are more aware of this than consumers. As a result, the holiday season is attracting a lot of attention and businesses are looking at various ways to attract customers. Economists call it the "consumption function" period, sociologists call it the "consumer behavior impulse" period and the business community calls it the "holiday effect". Making good use of this effect can bring objective economic benefits. Through the empirical study of the holiday effect of China's time-honored brand enterprises, I propose the following recommendations.

We will increase the R&D innovation of TCM, strengthen the supervision and management of TCM, promote the quality improvement of TCM, strengthen the comprehensive technical service guarantee and optimize the business environment of TCM to achieve industrial upgrading so that the development of the TCM concept stocks can break through the relevant bottlenecks. At the same time, TCM companies will give full play to their brand power and realize the brand effect, thereby promoting the development of TCM.

Increase the publicity for China's time-honored enterprises to increase their audience. Empirical tests have found that the Chinese New Year has a more significant festive effect on long-established companies. The reason for this is that the cultural element of the Spring Festival gives older companies greater exposure. Therefore, time-honored enterprises should increase their own publicity, enhance their own innovation, constantly increase their exposure, and at the same time expand their original consumer groups, break through the "old" as the core stereotype, and increase their own audience among young people. For example, Moutai's cross-border design of white wine ice cream is an excellent attempt to increase its exposure and continue to innovate.

**Author Contributions:** Conceptualization, L.W.; Methodology, Y.Z.; Software, X.Y.L.; Validation, Y.X.M.; Formal analysis, J.Y.X.; Resources, J.Y.X. & Y.X.M.; Data curation, X.Y.L. & J.Y.X.; Writing-original draft, X.Y.L. & Y.Z.; Writing-review&editing, X.Y.L. & L.W.; Visualization, X.Y.L. & L.W. All author have read and agreed to the published version of the manuscript.

**Funding:** This research received no external funding

**Conflicts of Interest:** The authors declare no conflicts of interest